\documentstyle[aps,12pt,epsf]{revtex}

\newcommand\pubdate{\today}
\newcommand\hepnumber{hep-ph/01xxx}

\def\csumb{Ottawa-Carleton Institute for Physics,
Department of Physics, Carleton University, Ottawa, Canada K1S 5B6
}

\def\Title#1{\begin{center} {\Large\bf #1 } \end{center}}
\def\Author#1{\begin{center}{ \sc #1} \end{center}}
\def\Address#1{\begin{center}{ \it #1} \end{center}}

\newcommand\pubblock{\rightline{\begin{tabular}{l} 
         \pubdate\\ \hepnumber \end{tabular}}}
\newenvironment{Abstract}{\begin{quotation}  }{\end{quotation}}

\def\lsim{\mathrel{\rlap{\lower4pt\hbox{\hskip1pt$\sim$}}
    \raise1pt\hbox{$<$}}}                
\def\gsim{\mathrel{\rlap{\lower4pt\hbox{\hskip1pt$\sim$}}
    \raise1pt\hbox{$>$}}}                

\def\a{\alpha}

\def\d{\delta}
\def\e{\epsilon}

\def\m{\mu}

\def\p{\pi}

\def\s{\sigma}

\def\G{\Gamma}

\def\bo{{\raise.15ex\hbox{\large$\Box$}}}         

\def\leftrightarrowfill{$\mathsurround=0pt \mathord\leftarrow \mkern-6mu
        \cleaders\hbox{$\mkern-2mu \mathord- \mkern-2mu$}\hfill
        \mkern-6mu \mathord\rightarrow$}       
\def\dvec#1{\vbox{\ialign{##\crcr
        \leftrightarrowfill\crcr\noalign{\kern-1pt\nointerlineskip}
        $\hfil\displaystyle{#1}\hfil$\crcr}}}           

\def\be{\begin{equation}}
\def\ee{\end{equation}}

\def\bex{\begin{displaymath}}
\def\eex{\end{displaymath}}

\def\bea{\begin{eqnarray}}
\def\eea{\end{eqnarray}}
\def\NO{\nonumber}

\def\beax{\begin{eqnarray*}}
\def\eeax{\end{eqnarray*}}

\begin{document}
\begin{titlepage}
\pubblock

\Title{ Complete next-to-leading order QCD corrections to charged Higgs boson
associated production with top quark at the CERN Large Hadron Collider}
\vfill
\Author{  Shou-hua Zhu }
\Address{\csumb}
\vfill
\begin{Abstract} 
The complete
next-to-leading order (NLO) QCD corrections to charged Higgs boson
associated production with top quark 
through $b g \rightarrow
tH^{-}$ at the  CERN Large Hadron Collider
are calculated in the minimal supersymmetric standard model (MSSM)
and two-Higgs-doublet model in the $\overline{MS}$ scheme. 
The NLO QCD corrections can reduce 
the scale dependence of the leading order (LO) cross section. The
K-factor (defined as the ratio of the NLO cross section to the
LO one) does not depend on $\tan\beta$ if the same quark running masses are used in the NLO and LO cross sections, and varies roughly from
$\sim 1.6$ to $\sim 1.8$ when charged Higgs boson mass 
increases from  $200$ GeV to $1000$ GeV.

\end{Abstract}
\vfill


\end{titlepage}

\eject \baselineskip=0.3in

\section{ Introduction}

The detection of the Higgs particles is
one of the most important objectives of the Large Hadron
Collider (LHC).
Charged Higgs bosons
are predicted in extended versions of the Standard model (SM),
like two-Higgs-doublet models (2HDM) and
the Minimal Supersymmetric Standard Model (MSSM).
Discovery of such an additional charged Higgs boson will
immediately indicate physics beyond the SM,
unlike the case of the neutral Higgs boson.
Hence, there is strong theoretical and experimental motivation for
exploring the mechanisms of the charged Higgs boson production.

The charged Higgs boson $H^\pm$ could appear as
the decay product of primarily produced top quarks
if the mass of $H^\pm$ is smaller than $m_t - m_b$.
For heavier $H^\pm$, the direct $H^\pm$ production mechanisms
at hadron colliders have been extensively investigated. 
At the LHC, the primary charged Higgs boson production
channel is 
$g b \to H^- t$\cite{Jin:2000vx} \footnote{
See Ref. \cite{Hollik} for the discussion on 
other charged Higgs boson production 
mechanisms.}. 
The study \cite{Roy:1999xw,Odagiri:1999xz} shows that  
this production mechanism can be used to explore 
the parameter space of MSSM
for $m_{H^{\pm}}$ up to 1 TeV and  $\tan\beta $ down to at least
$\sim 3$, and potentially to $\sim 1.5$.  Therefore, it is 
necessary to calculate and implement
also the loop contributions to $g b \rightarrow  H^- t$
for more accurate theoretical predictions.

The production mechanism of the heavy charged Higgs boson with
heavy top quark has been studied two decades ago \cite{Collins:mp}. In order
to resum the possible large terms like $\log(Q^2/m_b^2)$, the bottom
quark parton distribution function (PDF) is introduced. Although, 
there are some
doubts about the bottom parton description \cite{SpiraSUSY02}, the detailed
study on the $P_{T,b}$ distribution \cite{Tilman} 
argues that the bottom 
parton description is reliable for the inclusive $tH^-$ production
process at LHC. 

In earlier literature, 
the contribution of the initial-gluon process
$gg \rightarrow  H^- t \bar b$,
which is only part of the next-to-leading order (NLO) QCD corrections
to $g b \rightarrow  H^- t$,
has been calculated \cite{Borzumati:1999th}. The supersymmetric
electroweak corrections arising from the quantum effects which are induced
by potentially large Yukawa couplings from the Higgs sector and
the chargino-top(bottom)-sbottom(stop) couplings, neutralino-
top(bottom)-stop(sbottom) couplings and charged Higgs-stop-sbottom
couplings are also studied \cite{Jin:2000vx,Jin:1999tw},
which can give rise to a 15\% reduction of the lowest-order result.
In Ref. \cite{Belyaev:2001qm}, the electro-weak corrections to the process
are also discussed. After the submission of this paper, another calculation on the 
QCD correction appeared \cite{Tilman}, in which the corresponding results
seemed compatible with ours. 
The study on the SUSY-QCD effect 
for this process is also done
\cite{Tilman,Gao:2002is}.
In this paper, we would present the detailed and complete NLO
 QCD corrections to
$g b \rightarrow H^- t$.
We should note here that the results presented
in this paper
are for the process $bg \rightarrow t H^-$; they are
the same for the charge conjugate process $\bar b g \rightarrow H^+ \bar t$.

The arrangement of this paper is as follows. Section~II contains the
analytic results, and in Section~III we present numerical examples
and discuss the implications of our results.
The lengthy expressions of the form factors are
collected in the Appendix.

\section{Analytic expressions}

Including the NLO QCD corrections,
the cross sections for $P P \rightarrow tH^- X$ at the CERN
LHC can be written as 
\bea
\sigma=\sigma^{LO}+
\sigma^{Vir}+\sigma^{Real},
\eea
where $\sigma^{LO}$ is the cross section at leading order (LO),
$\sigma^{Vir}$  and
$\sigma^{Real}$ are cross sections from NLO QCD corrections arising
from virtual and real processes.

\subsection{LO cross section }

The Feynman diagrams for the charged Higgs boson production via
$b(p_1) g(p_2)\rightarrow t(k_1) H^{-}(k_2)$ at the LO are shown in Fig.1.
The amplitudes are created by use of Feynarts \cite{Kublbeck:1990xc}
and are handled with the help of FeynCalc \cite{Mertig:1991an}.
As usual, we define the Mandelstam variables as
\begin{eqnarray}
 s =(p_1 +p_2)^2 =(k_1 +k_2)^2, \nonumber \\  t =(p_1 -k_1)^2
=(p_2 -k_2)^2, \nonumber \\  u =(p_1 -k_2)^2 =(p_2 -k_1)^2.
\end{eqnarray}

The amplitude at the LO could be written as
\bea
M_{LO}= \sum_{i=1}^{6} t_i [c_1 M_{2 i-1}+c_2 M_{2 i}], 
\label{loamp}
\eea
where the non-vanishing form factors are
\bea
t_2 &=& \frac{1}{m_t^2-u}-\frac{1}{s}, \NO \\
t_3 &=& \frac{2 }{m_t^2-u}, \NO \\
t_5 &=&-\frac{2 }{s}, 
\eea
where
$c_1 = \frac{g_w g_s \mu^{2 \e} m_b \tan\beta}{2 \sqrt{2} m_w}$ and 
$c_2 = \frac{g_w g_s \mu^{2 \e} m_t \cot\beta}{2 \sqrt{2} m_w}$.
In Eq. (\ref{loamp})
$M_{2 i}$ and $M_{2 i+1}$ (i=1-6) are the standard matrix elements which are defined as
\bea
 &&M_1= \bar u(k_1) \not\varepsilon(p_2) P_R u(p_1),\nonumber
\\&&M_2= \bar u(k_1) \not\varepsilon(p_2) P_L u(p_1),\nonumber
\\&&M_{3}= \bar u(k_1) \not p_2 \not\varepsilon(k_2) P_R u(p_1),\nonumber
\\&&M_{4}= \bar u(k_1) \not p_2 \not\varepsilon(k_2) P_L u(p_1), \nonumber
\\&&M_5= \bar u(k_1) P_R u(p_1)k_1\cdot\varepsilon(p_2),\nonumber
\\&&M_6= \bar u(k_1) P_L u(p_1)k_1\cdot\varepsilon(p_2),\nonumber
\\&&M_7= \bar u(k_1) \not p_2 P_R u(p_1)k_1\cdot\varepsilon(p_2),\nonumber
\\&&M_8= \bar u(k_1) \not p_2 P_L u(p_1)k_1\cdot\varepsilon(p_2),\nonumber
\\&&M_9= \bar u(k_1) P_R u(p_1)p_1\cdot\varepsilon(p_2),\nonumber
\\&&M_{10}= \bar u(k_1) P_L u(p_1)p_1\cdot\varepsilon(p_2),\nonumber
\\&&M_{11}= \bar u(k_1) \not p_2 P_R u(p_1)p_1\cdot\varepsilon(p_2),\nonumber
\\&&M_{12}= \bar u(k_1) \not p_2 P_L u(p_1)p_1\cdot\varepsilon(p_2),
\eea
where the color matrix $T^a$ has been suppressed.
In this paper, we perform the calculations in Feynman gauge and in $d$ 
time-space dimensions with $d =4-2 \e$. For simplicity, throughout
the paper we omit the
bottom quark dynamical mass but keep the mass-term
only in Yukawa couplings.  The discussion on the error induced 
by massless bottom quark approximation can be found in Ref. 
\cite{Campbell:2002zm}. 

The LO cross section can then be written as
\bea
\frac{d\s_{LO}}{dx_1 dx_2}=d\hat \sigma^0
G_{b/A}(x_1,\m_f)G_{g/B}(x_2,\m_f) +\left[ A \leftrightarrow B \right],
\eea
with
\bea
d\hat \sigma^0=\frac{1}{24} \frac{1}{4 (1-\e)} \frac{1}{2 s} |M_{LO}|^2
d\Phi_2
\eea
where the factor $\frac{1}{24}$ and  $\frac{1}{4 (1-\e)}$ are the color and
spin average, respectively,
and two-body phase space is
\bea
d\Phi_2=
 \frac{1}{8 \pi} \left(\frac{4 \pi}{s}\right)^\e
\frac{1}{\Gamma(1-\e)} \left[
\lambda\left(1, \frac{m_{H^\pm}^2}{s},\frac{m_{t}^2}{s}\right)
 \right]^{1/2-\e} v^{-\e} (1-v)^{-\e} d v,
\eea
with $v=\frac{1}{2} (1+\cos\theta)$, where $\theta$ is the center-of-mass
scattering angle between $p_1$ and $k_1$.
Here $\lambda$ is the two-body phase space function
\bea
\lambda(x,y,z)=x^2+y^2+z^2-2 x y- 2 x z-2 yz.
\eea 


\subsection{Virtual corrections}

The Feynman diagrams for the NLO virtual corrections are shown in Fig. 2.
The virtual diagrams  with
the self-energy insertions on the external legs 
are not shown.  They can be obtained by inserting diagram (a)-(d) of Fig. 2 
into the external legs
of the LO diagrams in Fig. 1.  
At the same time, the diagrams containing counter-terms can be
easily got by inserting corresponding counter-terms into
external legs, internal propagators and
vertex of the LO diagrams.
  
In order to remove the UV divergences, we have to renormalize the
strong coupling constant, the Yukawa coupling constants, the quark masses 
and the wave functions
of quarks and gluon. The strong
coupling constant and the gluon wave function are renormalized 
in the $\overline{MS}$ scheme as
\bea
\frac{\delta g_s}{g_s}&=&-\frac{\alpha_s}{8 \pi}
 \beta_0 \Delta, 
\nonumber \\
Z_g &=& -\frac{\alpha_s}{4 \pi} 
 (2 C_A-\beta_0) \Delta,
\eea
with $\Delta=\frac{1}{\epsilon}-\gamma_E+\log(4 \pi)$,
 $\beta_0=(11 C_A-2 n_f)/3$,
$C_A=3$ and $C_F=4/3$. Here, $n_f=6$ is the number of
fermions.

We renormalize wave function and mass of the bottom quark in the $\overline{MS}$ scheme
\bea
\frac{\delta {m_b}}{m_b}&=&-\frac{\alpha_s}{4 \pi} 3 C_F \Delta,
\nonumber \\
Z_b&=&-\frac{\alpha_s}{4 \pi} C_F \Delta.
\eea
At the same time, we will renormalize the
top quark mass in two schemes: on-mass-shell (OS) and
$\overline{MS}$. 
\bea
\frac{\delta m_t}{m_t} &=&-\frac{\alpha_s}{4 \pi} 3 C_F
\left[ \Delta+\frac{4}{3}-\log(m_t^2/\mu^2) \right],
\ \ \ \mbox{\rm in \ OS \ scheme}
\nonumber \\
\frac{\delta m_t}{m_t} &=&
-\frac{\alpha_s}{4 \pi} 3 C_F \Delta, \ \ \ \mbox{\rm in \ $\overline{MS}$
 \ scheme}.
\eea
Hereafter we will refer OS and $\overline{MS}$ schemes to
the different mass renormalization of the top quark.
The comparison between the results in these two schemes will be discussed
in the numerical section.
The wave function renormalization constant of top quark
is chosen the same with that of b quark
\bea
Z_t&=&-\frac{\alpha_s}{4 \pi} C_F \Delta.  
\eea 

The renormalized virtual amplitude can then be
written in the following way,
\begin{eqnarray}
\label{amplitude}
 M_{\rm ren}& =& M_{A}+M_{B},
\end{eqnarray}
where $M_{A}$ is the amplitude 
from the diagrams (e)-(o) of Fig. 2 and 
$M_{B}$ is the amplitude
from the diagrams which contain self-energy insertion on
the external legs or counter-terms.

The $M_{A}$ can be written as 
\bea
M_{A}=\sum_{i=e}^{o} M_{A}^i,
\eea
where $i$ represents the diagram index of Fig. 2. 
For each diagram $i$, we can generally write
the amplitude as
\bea
M_{A}^i= \sum_{j=1}^{6} f_j [c_1 M_{2 i-1}+c_2 M_{2 i}],
\label{formfactor}
\eea
where the non-vanishing form factors $f_j$ are given explicitly
in Appendix and the $M_j$ are the standard matrix elements given
in the previous subsection.

The $M_{B}$ can be written as
\bea
M_{B}= M_{B}^1+ M_{B}^2,
\eea
where
\bea
M_{B}^{1}= M_{LO}
\left\{ \frac{\delta g_s}{g_s}+
\frac{Z_t}{2}+\frac{Z_b}{2}+
\frac{Z_g}{2}+
\frac{\alpha_s}{ 8 \pi} \left[
-3 \Delta+\frac{14}{3} \log(m_t^2/\mu^2)-
\frac{16}{3}  \right] \right\},
\eea
\bea
M_{B}^2=\sum_{i=1}^{6} [c_1 f_s^{2 i-1} M_{2 i-1}+c_2 
f_s^{2 i} M_{2 i}].
\eea
Here
the non-vanishing form factors $f_s^i$ $(i=1-12)$ are
\bea
f_s^1 &=& f_s^2= \frac{\delta m_t}{m_t^2-u},
\nonumber \\
f_s^3 &=&  
\frac{-2 m_t \delta m_t}{(m_t^2-u)^2}+ 
\frac{\delta m_b}{m_b} \left[ \frac{1}{m_t^2-u}-\frac{1}{s} \right],
\nonumber \\
f_s^4 &=&  
\frac{-2 m_t \delta m_t}{(m_t^2-u)^2}+ 
\frac{\delta m_t}{m_t} \left[ \frac{1}{m_t^2-u}-\frac{1}{s} \right],
\nonumber \\
f_s^5 &=&
2 \left[ \frac{-2 m_t \delta m_t}{(m_t^2-u)^2}+ 
\frac{\delta m_b}{m_b} \frac{1}{m_t^2-u}\right],
\nonumber \\
f_s^6 &=&
2 \left[ \frac{-2 m_t \delta m_t}{(m_t^2-u)^2}+
\frac{\delta m_t}{m_t} \frac{1}{m_t^2-u}\right],
\nonumber \\
f_s^9 &=&
-\frac{2}{s} \frac{\delta m_b}{m_b},
\nonumber \\
f_s^{10} &=&
-\frac{2}{s} \frac{\delta m_t}{m_t}.
\label{conter-factor}
\eea

After squaring the renormalized amplitude and performing the spin and color
summations, the partonic
cross section with virtual corrections
can be written as
\bea
\frac{d \sigma^{Vir}}{dx_1 dx_2}= 2\, {\rm Re} \left[ \overline{\sum}\, 
(M^+_{\rm ren} M_{LO}) \right]
d \Phi_2 G_{b/A}(x_1,\m_f)G_{g/B}(x_2,\m_f)+ \left[ A \leftrightarrow B \right],
\eea
where $A$, $B$ denote the incoming hadrons.

After the renormalization procedure described above, 
$d \sigma^{Vir}$ is UV-finite. Nevertheless, it contains still the
soft and collinear divergences.
The soft divergences will cancel against
the contributions from real-gluon radiation [see Eq. (\ref{eq:soft})]. 
The remaining collinear divergences in the real-gluon-emission processes
will be removed by the redefinition of the parton distribution functions (PDF)
(mass factorization) [see Eq. (\ref{eq:collinear})]. 

\subsection{Real corrections}

There are three kinds of real corrections to the processes
$bg \rightarrow t H^-$: gluon-radiation
[$bg \rightarrow t H^- g$], initial-gluon [
$gg \rightarrow t H^- \bar b$ ] and initial-active-quark
 [$bq (\bar q)
\rightarrow t H^- q (\bar q)$ and
$q \bar q \rightarrow t H^- \bar b$, where $q$ stand for the active quarks
which are treated as light for PDF evolution, in practice the light quarks
other than $u, d, s$ can be omitted due to the low luminosity]. 
All real corrections are related to the 
$2 \rightarrow 3$ processes. 

In this paper, the 
$2 \rightarrow 3$ processes
have been treated using the two
cut-off phase space slicing method (TCPSSM)
\cite{Harris:2001sx}. The method is briefly described
in the following. 
Two small artificial constants $\delta_s$, $\delta_c$ are
introduced, and
the three-body
phase space can firstly be divided into soft and hard regions
according to whether the gluon energy is less
than $\delta_s \sqrt{s}/2$.  
Secondly the hard region
is further divided into hard collinear and hard non-collinear
regions according to whether the magnitude of $p_i . p_j$ is
less than $\delta_c s/2$ ($p_i, p_j$ are the possible collinear 
momenta).
In the soft and collinear regions, the phase space integration
can be performed analytically in $d=4-2\epsilon$ dimensions. 
At the same time, in the hard non-collinear region, 
the phase space integration
can be calculated in four dimensions by standard Monte Carlo 
packages because
the integration contains no divergences. Obviously, the final 
physical results should be independent on these artificial
parameters $\delta_s$ and $\delta_c$, which offers a crucial
way to check our results. Therefore, the real corrections can
be written technically as, according to the phase space slicing,
\bea
 \sigma^{Real}=  \sigma^{S} +\overline{ \sigma^{Coll}}
+ \sigma^{fin}
\eea 
with
\bea
\overline{ \sigma^{Coll}}= \sigma^{Coll}+ \sigma^{fac},
\eea
where $ \sigma^{S}$, $ \sigma^{Coll}$ and $ \sigma^{fin}$
are cross sections for the 
direct calculations of the $2\rightarrow 3$ processes
in soft, hard collinear and hard non-collinear regions, and the 
$\sigma^{fac}$ is the counter-term from factorization procedure.
After adding virtual contributions and $ \sigma^{S}$, the double
poles are canceled, and the remaining singularities together
with $\sigma^{Coll}$ are canceled by $\sigma^{fac}$. In the following,
$ \sigma^{S}$, $\overline{ \sigma^{Coll}}$ and
$\sigma^{fin}$ will be presented respectively.

\subsubsection{Cross section in soft region}

The Feynman diagrams of the gluon-radiation process $bg \rightarrow
t H^- g$ are shown
in Fig. 3. Diagrams except (c) and (d) contribute to the cross section
in the soft region.  
We may write the cross section as
\bea
\frac{d\s^{S}}{dx_1 dx_2} &=& \hat{\s}^0_S 
 G_{b/A}(x_1,\m_f)G_{g/B}(x_2,\m_f)+\left[ A \leftrightarrow B \right], 
\nonumber \\
\hat{\s}^0_S &=& \hat{\s}^0 \left[ \frac{\a_s}{2\p}
\frac{\G(1-\e)}{\G(1-2\e)} \left( \frac{4\p\mu^2}{s} \right)^\e \right]
\left( \frac{A_2^s}{\e^2}+\frac{A_1^s}{\e}+A_0^s \right),
\label{eq:soft}
\eea
where
\bea
A_2^s&=&-\frac{4}{3} \frac{m_t^2-t}{(E_1-\beta \cos\theta) s}+
12 \frac{m_t^2-u}{(E_1+\beta \cos\theta) s}+12, \NO \\ \NO
A_1^s&=& \frac{16}{3}  \frac{4 m_t^2}{(E_1^2-\beta^2) s}+
\frac{4}{3} \frac{m_t^2-t}{(E_1-\beta \cos\theta) s} (C_1-
2 \log\frac{2}{\delta_s \sqrt{s}}) \\ \NO
&&+12 \frac{m_t^2-u}{(E_1+\beta \cos\theta) s} (C_2-2 
\log\frac{2}{\delta_s \sqrt{s}})+24  \log\frac{2}{\delta_s \sqrt{s}}-
\log\frac{4}{s} A_2^s,\\
\NO
A_0^s &=& \frac{1}{2} \log^2\frac{4}{s}  A_2^s-\log\frac{4}{s} A_1^s+
\frac{16}{3}  \frac{4 m_t^2}{(E_1^2-\beta^2) s} (
2 \log\frac{2}{\delta_s \sqrt{s}}+\frac{E_1}{\beta} \log\frac{E_1+\beta}{
E_1-\beta}) \\
&&+
\frac{4}{3}  \frac{m_t^2-t}{(E_1-\beta \cos\theta) s} (C_3-
2 \log^2\frac{2}{\delta_s \sqrt{s}}+2 C_1 \log\frac{2}{\delta_s \sqrt{s}})
\NO \\
&&-12 \frac{m_t^2-u}{(E_1+\beta \cos\theta) s} (C_4-
2\log^2\frac{2}{\delta_s \sqrt{s}}+2 C_2 \log\frac{2}{\delta_s \sqrt{s}})
 +24 \log^2\frac{2}{\delta_s \sqrt{s}}, \NO \\
C_1&=& \log\frac{(E_1-\beta \cos\theta)^2}{E_1^2-\beta^2},\NO \\
C_2&=& \log\frac{(E_1+\beta \cos\theta)^2}{E_1^2-\beta^2},\NO \\
C_3&=& -\log^2\frac{E_1-\beta }{E_1-\beta  \cos\theta} +
\frac{1}{2}\log^2\frac{E_1+\beta }{E_1-\beta}-
2 li_2(-\frac{-\beta \cos\theta+ \beta}{E_1-\beta})+
2 li_2(-\frac{\beta \cos\theta+\beta}{E_1-\beta \cos\theta}) \NO,\\ 
C_4&=& -\log^2\frac{E_1-\beta }{E_1+\beta  \cos\theta} +
\frac{1}{2}\log^2\frac{E_1+\beta }{E_1-\beta}-
2 li_2(-\frac{\beta \cos\theta+ \beta}{E_1-\beta})+
2 li_2(-\frac{-\beta \cos\theta+\beta}{E_1+\beta \cos\theta}), \NO \\
\beta &=& \lambda^{\frac{1}{2}}(1,m_t^2/s,m_{H^\pm}^2/s), \NO \\
E_1 &=& \sqrt{\beta^2+\frac{2 m_t^2}{s}},   
\eea
with $\cos\theta$ defined in II-A.

\subsubsection{Cross section in hard collinear region}

As discussed above, the real corrections in collinear region
$\s^{Coll}$
contain divergences because the partons are massless. 
In order to remove these kinds of collinear singularities,
we introduce  scale dependent parton distribution functions
in $\overline{\rm MS}$ convention
\bea
G_{c/P}(x,\mu_f)=G_{c/P}(x)-
\frac{1}{\epsilon} \left[\frac{\alpha_s}{2\pi}
\frac{\Gamma(1-\epsilon)}{\Gamma(1-2\epsilon)}
\left(\frac{4\pi\mu^2}{\mu_f^2}\right)^\epsilon \right]
\int^1_x \frac{dz}{z} P_{cc^\prime}(z) G_{c^\prime/P}(x/z),
\label{factorization}
\eea
where $P_{cc^\prime}(z)$ are splitting functions and 
$c [c^\prime]$ represent $b$ or $g$ [partons which can
split into $c$].
After substituting the $G_{c/P}(x)$ in the lowest order
expressions by $G_{c/P}(x,\mu_f)$, we can obtain $\s^{fac}$.
Adding $\s^{Coll}$ and $\s^{fac}$, we can get the
final results for $\overline{\s^{Coll}}$
\begin{eqnarray}
\frac{\overline{d\s^{Coll}}}{dx_1 dx_2} &=& 
\left[ \frac{\alpha_s}{2\pi} \frac{\G(1-\e)}{\G(1-2\e)}
\left(\frac{4 \pi \mu^2}{s}\right)^{\e}\right] 
\left\{
 G_{b/A}(x_1,\m_f)G_{g/B}(x_2,\m_f) \right. \NO \\
&&\times \left[ 
 \frac{A_1^{sc}(b\rightarrow bg)}{\e}+
 \frac{A_1^{sc}(g\rightarrow gg)}{\e}
+ A_0^{sc}(b\rightarrow bg)+
A_0^{sc}(g\rightarrow gg) \right]
 \NO \\
&&+  \left.  
G_{b/A}(x_1,\m_f) \widetilde{G}_{g/B}(x_2,\m_f)
+ \widetilde{G}_{b/A}(x_1,\m_f) G_{g/B}(x_2,\m_f)
\right\} \hat\s_0 \NO \\
&&+ \left[ A \leftrightarrow B \right],
\label{eq:collinear}
\eea
where \cite{Harris:2001sx}
\bea
A_0^{sc} &=& A_1^{sc} \ln \left( \frac{s}{\mu_f^2} \right)
 \\
A_1^{sc}(b\rightarrow bg) &=& C_F(2 \ln \d_s + 3/2 )  \\
A_1^{sc}(g\rightarrow gg) &=& 2N \ln \d_s + (11N-2 n_f)/6 \, 
\eea
with $n_f=5$ is the number of massless fermions.
Here
\be
\widetilde{G}_{c/B,A}(x,\m_f) = \sum_{c'}  \int_x^{1-\d_s\d_{cc'}} \frac{dy}{y}
                       G_{c'/B,A}(x/y,\m_f) \widetilde{P}_{cc'}(y)
\ee
with
\be
\widetilde{P}_{ij}(y) = P_{ij}(y)\ln\left(\d_c\frac{1-y}{y}
\frac{s}{\m_f^2}\right) - P_{ij}^{\prime}(y) \,,
\ee
where
\bea
P_{qq}(z) &=& C_F \frac{1+z^2}{1-z} \\
P_{qq}^{\prime}(z) &=& -C_F(1-z) \\
P_{gq}(z) &=& C_F \frac{1+(1-z)^2}{z} \\
P_{gq}^{\prime}(z) &=& -C_Fz \\
P_{gg}(z) &=& 2N\left[ \frac{z}{1-z}+\frac{1-z}{z}+z(1-z)\right] \\
P_{gg}^{\prime}(z) &=& 0 \\
P_{qg}(z) &=& \frac{1}{2} \left[ z^2+(1-z)^2 \right] \\
P_{qg}^{\prime}(z) &=& -z(1-z) \, ,
\eea
with $N=3$.

\subsubsection{Cross section in hard non-collinear region}

As described above, the cross section in hard non-collinear region
$ \sigma^{fin}$
can be easily obtained by Monte Carlo phase space integration
in four dimension. It can be written as
\bea
\frac{d \sigma^{fin}}{dx_1 dx_2} =
\sum_{c,c^\prime} G_{c/A}(x_1,\mu_f) G_{c^\prime/B}(x_2,\mu_f)
|c c^\prime \rightarrow t H^- X|^2\ d \Phi_3 +
\left[ A \leftrightarrow B \right],
\eea
where $c,c^\prime$ run through gluon and light quarks and 
the three-body phase space $\Phi_3$ is within the hard
non-collinear region.
In this paper, all Monte Carlo phase space integrations are performed
by package BASES \cite{Bases}.

\section{Numerical results and discussion}

Our numerical results are obtained using CTEQ5M (CTEQ5L) PDF
\cite{Lai:2000wy} 
and $2$-loop ($1$-loop) evolution of $\alpha_s(\mu)$
for NLO (LO) cross section calculations with
$\Lambda^{(5)}= 226$ (146) MeV. As we have mentioned in the
second section, the dynamical mass of the bottom quark has been set
to zero, which means that we treat the bottom quark as usual massless 
partons. 
In the $\overline{MS}$ scheme,
$2$-loop evolution of the 
quark masses 
is adopted, and 
the pole masses of bottom and top quarks are taken as 4.7 GeV and 175 GeV. 
In the OS scheme (defined in this paper), the top quark 
mass is equal to the pole mass and the bottom quark mass
is the same with that in the $\overline{MS}$ scheme.
For simplicity, 
the renormalization and factorization scales
are taken to be the same.

In Fig.~4, we show the cross sections as a function of $\delta_s$ with $\delta_c=\delta_s/50$ in the $\overline{MS}$ scheme. It is obviously true that, as stated in II.C, the physical cross sections are independent on the artificial parameters $\delta_s$ and $\delta_c$.

\subsection{Theoretical uncertainties}

There are many theoretical uncertainties in the calculation of the cross sections, for examples
unknown scales of renormalization and factorization as well as the different choice
of renormalization schemes. We can define two quantities $\delta$ and $\Delta$ to measure
such kinds of uncertainties as
\bea
\delta(\mu)=\frac{\sigma_{OS}(\mu)-\sigma_{\overline{MS}}(\mu)}
{\sigma_{OS}(\mu)+\sigma_{\overline{MS}}(\mu)},  \\
\Delta (\mu)= \frac{\sigma (2\mu)-\sigma(\mu/2)}{\sigma(2\mu)+\sigma(\mu/2)}.
\eea

It is known that in perturbative calculations, the physical results are
independent on the renormaliztion schemes provided that we can expand
the perturbative series to infinity. Otherwise, the results do depend
on the renormalization schemes.
Therefore $\delta$ is due to the unknown parts of higher order effect
of the perturbative expansion series,
and it can be enhanced by the large term $\log(\mu^2/m_b^2)$, 
especially for large $\tan\beta$ case, where the yukawa coupling containing $m_b$ dominate the contribution. For such case, $\delta$ can be the measure for the entire 
theoretical uncertainties.
Explicitly, for large $\tan\beta$ limit  
\begin{eqnarray}
\sigma^{OS}_{LO} & \propto & m_b^2,
\nonumber \\
\sigma^{\overline{MS}}_{LO} & \propto & \overline{m}_b(\mu)^2.
\end{eqnarray}
Therefore
\bea
\delta_{LO}= \frac{m_b^2-\overline{m}_b(\mu)^2}{
m_b^2+\overline{m}_b(\mu)^2} \simeq \frac{A}{2}
\eea
with (at one-loop order)
\begin{eqnarray}
 A = 2 \frac{\alpha_s}{\pi} (\log\frac{\mu^2}{m_b^2}+\frac{4}{3})\, ,
\label{quarkmass}
\end{eqnarray}
which is the quantity entering also the relation between
the $\overline{MS}$ quark mass
and the corresponding pole mass,
\begin{eqnarray}
 m(\mu)^2 =m^2\,
 [1-A] \, .
\end{eqnarray}
For $\mu=\mu_0 \equiv m_{H^\pm}+m_t$, $\delta_{LO}\simeq 30\%$
if we choose $ m_{H^\pm}=200$ GeV. 

At NLO, we can write
\begin{eqnarray}
\sigma^{OS}_{NLO} & \propto & m_b^2\, (1-A+B),  \nonumber \\
\sigma^{\overline{MS}}_{NLO} & \propto & \overline{m}_b(\mu)^2\,  (1+B),
\end{eqnarray}
where $B$ is the $O(\alpha_s)$ radiative correction to the
LO cross section in the $\overline{MS}$ scheme.
Hence, one finds
\begin{eqnarray}
\delta_{NLO} &\simeq &  \frac{A B}{2} \, .
\end{eqnarray}
For the same $\mu$ in the last paragraph,
we have $B \simeq 0.3$,
and $\delta_{NLO} \simeq 10\%$.

For the small $\tan\beta$ case, where the yukawa coupling containing top quark mass 
dominates the contribution, $\delta$ can be obtained in the similar method as above. 
For $\mu=\mu_0=375$ GeV, 
$\delta_{LO} \simeq 10\%$ and $\delta_{NLO} \simeq 3\%$. For the intermediate value of $\tan\beta$, $\delta$ lies between the large and small $\tan\beta$ case.

From above discussion on $\delta$, we can see that $\delta$ at NLO is not large for the small
$\tan\beta$ case. Therefore the theoretical uncertainties 
from varying the renormalization and factorization scales might be more important. 
In Fig.~5, we show in (a): the LO and NLO cross sections
in OS and $\overline{MS}$ schemes and in (b):
the relative deviation $\delta$
as a function of renormalization and factorization scales $\mu/\mu_0$
for $m_{H^\pm}=200$ GeV and $\tan\beta=2$. 
From the figure we can see that the direct calculation confirms the above estimation for $\delta$. Furthermore, the NLO results reduce
the scale-dependence in both schemes.  From the figure we can also calculate
$\Delta_{LO, NLO}\simeq 6\%, 0.2\%$ 
in OS scheme and  $\Delta_{LO, NLO}\simeq 17\%, 5\%$ in $\overline{MS}$ scheme for 
$\mu=\mu_0$ GeV. 

From the discussion on $\delta$ and $\Delta$, which act as the measures of the theoretical
uncertainties, we can see that the theoretical uncertainties of the
cross sections at NLO are much smaller than that at LO. 

\subsection{Numerical results in the $\overline{MS}$ scheme}

In this section, we will give the numerical results in the 
$\overline{MS}$ scheme, in which the large terms like
$\alpha_s \log(m_b^2/\mu^2)$ have been resummed into
the running of the b quark mass. Therefore it gives
usually more stable results with respect to missing higher-order terms.

In Fig.~6 we show the K-factor, which is defined as
\bea
 K=\frac{\sigma_{NLO}}{\sigma_{LO}}, 
\label{kdef1}
\eea
as a function of the charged Higgs mass 
with the renormalization and factorization 
scales $\mu=\mu_0$.
In the $\overline{MS}$ scheme, if the identical Yukawa couplings
are used at LO and NLO calculations, the K-factor does not
depend on $\tan\beta$, which is not true in  
the OS scheme because the different 
mass renormalization constants of top and bottom quarks
spoil it.  
In Fig.~6,
the different contributions to K-factor 
from improved Born (which is equal to 1 if the difference between the LO 
and NLO PDF and $\alpha_s(\mu)$ is omitted), virtual+gluon-radiation,
initial-gluon, $bq$ ($\bar q$) ($q$ stand for the light quarks)
and $q \bar q$ are also shown.   
From the figure, we can see that K-factor from improved Born contribution
is around $1.2 \sim 1.3$. K-factor from
virtual+gluon-radiation contribution
is from $0.7$ to $0.9$ when the charged Higgs boson mass
varies from 200 GeV to 1000 GeV. The initial-gluon and 
$bq (\bar q)$ contributions to the K-factor
are negative, and they vary from $\sim -27\%$ to $\sim -24 \%$ 
and  $\sim -5\%$ to $\sim -14 \%$ respectively.
The $q \bar q$ contribution to the K-factor
can be neglected, the magnitude of which is
smaller than  $3\%$ for all charged Higgs boson mass.
Adding all the contributions, we can see that the K-factor varies
from $\sim 1.6$ to $\sim 1.8$ when charged Higgs mass
increases from 200 GeV to 1000 GeV.

In Eq. (\ref{kdef1}), if we replace the two-loop evolution
quark masses by one-loop evolution one in the calculation of
the LO cross section, in the $\overline{MS}$ scheme,  the $K^\prime$ can be expressed as
\bea
K^\prime= K \frac{\overline{m}_{t,(2)}^2 +
\overline{m}_{b,(2)}^2 \tan^4\beta}{
\overline{m}_{t,(1)}^2 +
\overline{m}_{b,(1)}^2 \tan^4\beta},\label{kfac2}
\eea
where the subscripts of quark masses are which kind of quark mass evolutions
are used. In Fig.~7, the $K^\prime$ is shown as a function of charged Higgs boson mass with $\tan\beta=2, 5, 10, 30$, respectively. The dependence on
$\tan\beta$ can be explained by Eq. (\ref{kfac2}). \\

To summarize, 
the next-to-leading order QCD corrections to charged Higgs boson
associated production with top quark
through $b g \rightarrow
tH^{-}$ at the  CERN Large Hadron Collider
are calculated in the minimal supersymmetric standard model 
and two-Higgs-doublet model in the $\overline{MS}$ scheme.
It should be noted that in MSSM, the SUSY-QCD corrections arising
from the virtual gluino and squarks should be also included. For
some specific parameters, the SUSY-QCD can be significant
\cite{Tilman,Gao:2002is}. From the calculations,
we can see that the NLO QCD corrections can reduce
the scale dependence of the LO cross section. The
K-factor
does not depend on $\tan\beta$ if the same quark running masses are used in the NLO and LO
cross sections, and varies
 from $\sim 1.6$ to $\sim 1.8$ when charged Higgs mass 
increases from  $200$ GeV  to $ 1000$ GeV.

\section{Acknowledgement}
The author would like to thank Prof. W. Hollik,  Prof. C.S. Li,
Prof. C.P. Yuan
and Dr. J. Guasch for stimulating discussions.
This work was supported in part by the Nature Sciences and Engineering Research Council of Canada, the Alexander von Humboldt
Foundation and
National Nature Science Foundation of China.
Parts of the calculations have been performed on the QCM
cluster at the University of Karlsruhe, supported by the
DFG-Forschergruppe ''Quantenfeldtheorie, Computeralgebra und
Monte-Carlo-Simulation''.

\section{Appendix}

In this appendix, we will give the non-vanishing form-factors in
Eq. (\ref{formfactor}). For completeness, we give firstly the
definition of loop integrals and its Lorentz decomposition:
\bea
B_0(p_1^2,m_0^2,m_1^2)&=&\frac{(2\pi\mu)^{4-d}}{i \pi^2}
\int d^dq \frac{1}{Q_0 Q_1 }, 
\eea
\bea
&&C_{0;\mu;\mu\nu} (p_1^2,p_{12},p_2^2,m_0^2, m_1^2,m_2^2)
= \frac{(2\pi\mu)^{4-d}}{i \pi^2}
\int d^dq \frac{1;q_\mu;q_\mu q_\nu}{
Q_0 Q_1 Q_2},
\eea
\bea
&&D_{0;\mu;\mu\nu}(p_1^2,p_{12},p_{32},p_3^2,p_2^2,
p_{31},m_0^2, m_1^2,m_2^2,m_3^2)
= \frac{(2\pi\mu)^{4-d}}{i \pi^2}
\int d^dq \frac{1;q_\mu;q_\mu q_\nu}{
Q_0 Q_1 Q_2 Q_3}
\eea
with
\bea
Q_0=q^2-m_0^2+i\epsilon, \ \ \ Q_i=(q+p_i)^2-m_i^2+i\epsilon,
\ \ \ p_{ij}=(p_i-p_j)^2
\eea
and
\bea
C_{\mu}&=&p_{1\mu} C_1+p_{2\mu} C_2 \NO \\
C_{\mu\nu}&=&g_{\mu\nu} C_{00}+p_{1\mu}p_{1\nu} C_{11}
+(p_{1\mu}p_{2\nu}+p_{2\mu}p_{1\nu}) C_{12}+
p_{2\mu}p_{2\nu} C_{22} \NO \\
D_{\mu}&=&p_{1\mu} D_1+p_{2\mu} D_2+
p_{3\mu} D_3 \NO\\
D_{\mu\nu}&=&g_{\mu\nu} D_{00}+ p_{1\mu}p_{1\nu} D_{11}
+(p_{1\mu}p_{2\nu}+p_{2\mu}p_{1\nu}) D_{12}+
(p_{1\mu}p_{3\nu}+p_{3\mu}p_{1\nu}) D_{13} \NO\\
&& + p_{2\mu}p_{2\nu} D_{22}
+(p_{2\mu}p_{3\nu}+p_{3\mu}p_{2\nu}) D_{23}
+ p_{3\mu}p_{3\nu} D_{33}.
\label{eq:decomposition}
\eea

For simplicity, we define  abbreviation
for $B_0^i (i=1-7)$, $C_{x}^i
(i=1-8)$ [$x$ stands for the subscript defined in Eq.
(\ref{eq:decomposition})], $D_{0}^i (i=1-3)$ as \bea B_0^{1}&=&B_0(0, 0, m_t^2), \NO \\
B_0^{2}&=&B_0(0, m_t^2,m_t^2),  \NO \\ B_0^{3}&=&B_0(m_{H^\pm}^2,
0,m_t^2),  \NO \\ B_0^{4}&=&B_0(m_{t}^2, 0,m_t^2),  \NO \\
B_0^{5}&=&B_0(s,0,0),  \NO \\ B_0^{6}&=&B_0(t,0,m_t^2),  \NO \\
B_0^{7}&=&B_0(u,0,m_t^2),  \NO \\ 
C_x^{1}&=&C_x(0,0,s,0,0,0),  \NO
\\ C_x^{2}&=&C_x(0,m_{H^\pm}^2,t,m_t^2,m_t^2,0),  \NO \\
C_x^{3}&=&C_x(m_{H^\pm}^2,0,t,m_t^2,0,0),  \NO \\
C_x^{4}&=&C_x(m_{H^\pm}^2,0,u,m_t^2,0,0),  \NO \\
C_x^{5}&=&C_x(m_{H^\pm}^2,m_t^2,s,0,m_t^2,0),  \NO \\
C_x^{6}&=&C_x(m_{t}^2,0,t,m_t^2,0,0),  \NO \\
C_x^{7}&=&C_x(m_{t}^2,0,u,0,m_t^2,m_t^2),  \NO \\
C_x^{8}&=&C_x(m_{t}^2,0,u,m_t^2,0,0),  \NO \\ D_0^{1}&=&
D_0(m_{H^\pm}^2,0,m_t^2,0,t,u, 0,m_t^2,m_t^2,0),  \NO \\
D_0^{2}&=& D_0(m_{H^\pm}^2,m_t^2,0,0,s,t, 0,m_t^2,0,0),  \NO \\
D_0^{3}&=& D_0(m_{H^\pm}^2,m_t^2,0,0,s,u, 0,m_t^2,0,0). \eea

For diagram (e) in Fig. 2, we can write the form factor as
 \bea f_i=\frac{C_A}{2}\frac{ g_s^3}{16 \pi^2 s}
g_i \eea \bea g_2&=& -2 B_0^{5}-s [C_0^{1}+3 (C_1^1+C_2^1)]+4 (-1+2
\e) C_{00}^1, \NO \\ g_5&=& 2[  B_0^{5}(1+ \e) +s (C_0^{1}+2 C_1^1-3
C_2^1)]. \eea

For diagram (f) in Fig. 2, we can write the form factor as 
\bea f_i=(C_F-C_A/2)\frac{ g_s^3}{8 \pi^2 s} g_i
\eea \bea g_2&=& -B_0^{5}-(1+\e) s (C_0^{1}+C_1^1+C_2^1)+2 (1-\e)
C_{00}^1, \NO \\ g_5&=& (1+\e) B_0^{5}+2 \e s C_1^1-2 (1+\e) s C_2^1.
\eea

For diagram (g) in Fig. 2, we can write the form factor as 
 \bea f_i=C_F\frac{ g_s^3}{8 \pi^2 s} g_i \eea \bea
g_1&=& m_t s (C_0^{5} (1-\e) +C_2^{5}-\e (C_1^{5}+C_2^{5})), \NO \\
g_2&=&\frac{g_5}{2}= C_0^{5} (m_{H^\pm}^2-2 m_t^2)-B_0^{4}-B_0^{5}
-m_t^2 (C_1^{5}+C_2^{5}) +\e (B_0^{3}+s C_2^{5}). \eea

For diagram (h) in Fig. 2, we can write the form factor as
\bea f_i=C_F\frac{ g_s^3}{8 \pi^2 (m_t^2-u)} g_i
\eea \bea g_1&=& m_t (-1+\e) (m_t^2-u) (C_0^{4}+C_1^{4}+C_2^{4}), \NO \\
g_2&=&\frac{g_3}{2}= -C_0^{4} [ m_{H^\pm}^2+(-2+\e) m_t^2-\e u]
+B_0^{7} \NO \\ &&+ m_t^2 (C_1^{4}+C_2^{4}) -\e [B_0^{3}+(m_t^2-u)
(C_1^{4}+C_2^{4})]. \eea

For diagram (i) in Fig. 2, we can write the form factor as
 \bea f_i= \frac{C_A}{2} \frac{  g_s^3}{16 \pi^2
(m_t^2-u)} g_i \eea \bea g_1&=& 3 m_t (m_t^2-u) C_2^{8}, \NO \\ g_2&=&
2 B_0^{4}+2 B_0^{7}+ (m_t^2-u)(2 C_0^{8}+3 C_1^{8})+4(1-\e) C_{00}^{8},
\NO \\ g_3&=& 2 \{ \e -1+C_0^{8} (m_t^2-u) -(1+\e) B_0^{4}+2
B_0^{7}+ m_t^2 [ -5 C_1^{8}+4 C_2^{8}+2 (1-\e) C_{11}^{8}] \NO \\ && +u [ C_1^{8}
-4 C_2^{8}-2 (1-\e) C_{11}^{8}] \}, \NO \\ g_4&=& 4 m_t [C_1^{8} -2 C_2^{8}-
 (1-\e) C_{11}^{8}].
\eea

For diagram (j) in Fig. 2, we can write the form factor as
 \bea f_i=(C_F-C_A/2) \frac{g_s^3}{8 \pi^2
(m_t^2-u)} g_i \eea \bea g_1&=& m_t (m_t^2-u) [C_2^{7}+ \e
(C_0^{7}+C_2^{7})], \NO \\ g_2&=& 2 C_0^{7} m_t^2- (1+\e)
B_0^{2}+B_0^{4}+B_0^{7}+ (1+\e) (m_t^2-u) C_1^{7}+2 (-1+\e) C_{00}^{7},
\NO \\ g_3&=& -1-B_0^{4}+2 B_0^{7}+ 2 u C_{22}^{7}-2 m_t^2 ( 2
C_2^{7}+C_{22}^{7}) \NO \\ && -\e [ -1+B_0^{4}-2 (m_t^2-u) (C_2^{7}+ C_{22}^{7})],
\NO \\ g_4&=& -2 m_t [C_0^{7} \e+ (-1+2 \e) C_2^{7}+ (-1+\e) C_{22}^{7}].
\eea

For diagram (k) in Fig. 2, we can write the form factor as
 \bea f_i=C_F \frac{ g_s^3}{16 \pi^2 s} g_i \eea
\bea g_2&=&\frac{g_5}{2}= (1-\e) B_0^{5}. \eea

For diagram (l) in Fig. 2, we can write the form factor as
 \bea f_i=C_F \frac{ g_s^3}{16 \pi^2 (m_t^2-u)^2
u} g_i \eea \bea g_1&=& -m_t (m_t^2-u) \left\{ (\e-1) m_t^2
(B_0^{1}-B_0^{7})+ u (\e-3) B_0^{7} \right\}, \NO \\
g_2&=&\frac{g_3}{2}= (-1+\e) m_t^4 (B_0^{1}-B_0^{7})-(-1+\e) u^2
B_0^{7} \NO \\ &&+m_t^2 u \left\{(1-\e) [B_0^{1}-2 (1+B_0^{2}])]+2
(-3+\e) B_0^{7} \right\}. \eea

For diagram (m) in Fig. 2, we can write the form factor as
\bea f_i=\frac{C_A}{2}\frac{  g_s^3}{16 \pi^2} g_i
\eea \bea g_1&=& -m_t  [C_0^{8} +C_1^8+C_2^8- 2 (s-u+m_t^2)
(D_0^{3}+D_1+D_2+D_3) +4 D_{00}], \NO \\ g_2&=& -2
(C_0^{4}+C_0^{5})-2 D_0^{3} (m_t^2-u) + C_1^1+C_1^8+ 2 (t+2 u) (D_2+D_3)
\NO
\\ &&-m_{H^\pm}^2 (D_1+2 D_2+D_3)-m_t^2 (D_1+4 D_2+3 D_3)+4 \e
D_{00}, \NO \\ g_3&=& -2 \left\{ 2 [C_0^{4}-(1+\e)
C_0^{5}+C_0^{8}]-2 \e (C_1^5+C_2^5)+ C_1^8+C_2^8+2 D_0^{3} [-m_{H^\pm}^2+
(1-\e) m_t^2 \right. \NO \\ && +(1+\e) u] + 2[( 3-2 \e) m_t^2-t+2
u \e](D_1+D_2+D_3)
 + 2[ (1-\e) m_t^2+ u \e] (D_{11}
\NO \\ && \left. +2 D_{12}+2D_{13}+ D_{22}+2 D_{23}+D_{33})
\right\}, \NO \\ g_4&=& 4 m_t \left\{ (D_2+D_3+D_{11}+2
D_{12}+D_{13}+D_{22}+D_{23}) -\e [D_0^{3}+D_{11}+D_{22}+D_{33}
\right. \NO \\ &&\left. +2 (D_1 + D_2
+D_3+D_{12}+D_{13}+D_{23})]\right\}, \NO \\ g_5&=& 4 m_t \left\{
C_0^{1}+2\e C_1^5-C_1^1+ 2 (u-m_t^2) D_1+ [m_{H^\pm}^2-(7-2 \e) m_t^2+
2 (t+2 u)- 2 u\e] D_2 \right. \NO \\ &&\left. +
[-m_{H^\pm}^2-m_t^2+2 s] D_3+ [2 (\e-1) m_t^2- 2 u
\e](D_{12}+D_{22}+D_{23}) \right\},  \NO \\ g_{6}&=& -4 m_t
[D_0^{3}+D_1+(2-\e) D_2+D_3+D_{12}+D_{22}], \label{ffb1} \eea
where the variable of the D-function is the same with $D_0^3$.

For diagram (n) in Fig. 2, we can write the form factor as
 \bea f_i=(C_F-C_A/2)\frac{ g_s^3}{8 \pi^2} g_i
\eea \bea g_1&=& -m_t \left\{ D_0^{1} [s+\e(m_t^2-u)]+s (D_2 +D_3)
\right. \NO \\ &&\left.  -\e [C_2^6-m_t^2 (D_1+D_3)+u (D_2+D_3)] -2
D_{00} \right\}, \NO \\ g_2&=& -C_0^{4}-C_0^{7}+C_0^{2} (2+\e)
+D_0^{1} [ m_{H^\pm}^2+ \e m_t^2 -(2+\e) u] +D_1 [m_t^2 (1+\e)
 \NO \\
&&
+s- u(1+\e) ]
 +(D_2+D_3) [\e m_t^2 +s - u(1+\e) ]
-2 \e D_{00}, \NO \\ g_3&=& 2 \left\{ -C_0^{4} - C_0^{7} - 
(C_1^7+C_2^7)- \e (C_1^2+C_2^6) +2 m_t^2 D_2 - t (D_2+D_3) +m_t^2 (2 D_3+D_{22} \right.
\NO \\ &&\left. +2 D_{23}+D_{33}) - \e \left[C_2 +(m_t^2-u)
[D_2+D_3+D_{22}+2 D_{23}+D_{33}]\right] \right\}, \NO \\ g_4&=& 2
m_t \left\{ -D_2-D_3-D_{12}-D_{13}-D_{22}-2 D_{23}-D_{33} \right.
\NO \\ && \left. +\e [D_0^1+2 D_2+2 D_3+D_{22}+2
D_{23}+D_{33}]\right\}, \NO \\ g_5&=& -2 \left\{ C_0^{6} (-1+\e)+
(4 m_t^2 -t-2 u) D_2-s D_3+ m_t^2 (D_{22}+D_{23}) \right. \NO \\
&&\left. + \e [2 C_1^6-C_1^2+C_2^6-(m_t^2-u) (D_3+D_{22}+D_{23})] \right\},
\NO \\ g_6&=& 2 m_t [D_0-(-3+\e) D_2 +D_{12}-(-1+\e)
(D_{22}+D_{23})], \eea
where the variable of the D-function is the same with $D_0^1$.

For diagram (o) in Fig. 2, we can write the form factor as
 \bea f_i=(C_F-C_A/2)\frac{ g_s^3}{8 \pi^2} g_i
\eea \bea g_1&=& -m_t [C_0^{5}-\e C_0^{6}+D_0^{2} (1+\e) s- \e
(C_1^6+ C_2^6)+ s (1+\e) (D_1 +D_2 +D_3) -2 D_{00}], \NO \\ g_2&=&
-C_0^{1}-C_0^{5}+C_0^{3} (2+\e) +D_0^{2} (-2 m_t^2+t)- m_t^2
(D_1+D_2) \NO
\\ && +[s (1+\e)-u] D_3 -2 \e D_{00}, \NO\\ g_3&=& 2 \left\{ C_0^{5}
(1+\e) - C_0^{6} \e+ D_0^{2} [ 2 m_{H^\pm}^2+ m_t^2 (4-\e)- s
(2-\e) -3 t- u ( 2-\e)] \right. \NO \\ && +\e (C_1^3+C_1^5-C_1^6+C_2^5-C_2^6) +(D_1+D_2+D_3)
[m_{H^\pm}^2 (-3+\e)-m_t^2 \e+ 3 s +t (2-\e)
 \NO \\
&&
+u (3+\e)] 
\left.
 +(D_{11}+2 D_{12}+2 D_{13}+D_{22}+2 D_{23}+D_{33}) [m_t^2 (1-\e) +\e u]
\right\}, \NO \\ g_4&=& 2 m_t \left\{-D_3-D_{13}-D_{23}-D_{33}+ \e
(D_0+2 D_1+2 D_2 +2 D_3 +D_{11} \right. \NO \\ &&\left. + 2
D_{12}+2 D_{13}+D_{22}+2 D_{23}+D_{33})\right\}, \NO \\ g_5&=& -2
\left\{C_0^{6}-D_0^{2} ( m_{H^\pm}^2-2 m_t^2)-C_1^1 +\e(C_1^3+C_2^5) 
+D_1 [-t (1+\e)-2 u]+ \right. \NO \\ &&(D_2+D_3) [\e
m_{H^\pm}^2-s (1+\e)]+ D_3 [-t (1+\e) -u] +(D_{11}+D_{12}+2
D_{13}+D_{23}+D_{33}) u \e \NO \\ && \left. +m_t^2 [5 D_1+D_2+4
D_3 - (-1+\e) (D_{11}+D_{12}+2 D_{13}+D_{23}+D_{33})]\right\}, \NO
\\ g_6&=& 2 m_t\left\{ D_0^{2}-(-2+\e) (D_1+ D_3)+D_{13}+D_{33}- \e
(D_{11}+D_{12}+2 D_{13}+D_{23}+D_{33})\right\}, \eea
where the variable of the D-function is the same with $D_0^2$.

By decomposition, the loop integrals in above form-factors 
can be calculated by the limit number of
scalar integrals $B_0, C_0$ and $D_0$. The scalar integrals
$C_0$ and $D_0$
are UV finite, however some of them
contain soft and collinear divergences. Because
the finite scalar integrals could be calculated
by numerical method \cite{vanOldenborgh:1990wn}, only the
divergent ones are presented explicitly in this paper. It should be noted that 
only real part of
the integrals, which is relevant to our results, is given.

$D_0$ and $C_0$ scalar integrates
 could be generally written as \bea D_0 &=&(4 \pi
\mu^2)^\e \Gamma(1+\e) (\frac{d_2}{\e^2}+
\frac{d_1}{\e}+d_0),  
\NO \\
C_0 &=&-(4 \pi \mu^2)^\e
\Gamma(1+\e) (\frac{c_2}{\e^2}+ \frac{c_1}{\e}+c_0). \eea 

The coefficients $d_2, d_1$ and $d_0$ of $D_0^1$ are
\bea d_2 &=&
\frac{1}{2\,\left(  {{m_t}}^2 -t \right) \,
    \left(  {{m_t}}^2 -u\right) },
\eea
\bea
d_1 &=&
\frac{1}{\left(  {{m_t}}^2 -t \right) \,
    \left(  {{m_t}}^2-u \right) } \log\frac{m_t (m_t^2-m_{H^\pm}^2)}{
({{m_t}}^2 -t )({{m_t}}^2 -u )},
\eea
\bea
d_0 &=&
\frac{1}{2\,
    \left(  {{m_t}}^2 -t\right) \,
    \left(  {{m_t}}^2 -u \right) }
\left\{ \pi^2 \left[\theta(m_{H^\pm}-m_t)-\frac{1}{3}\right]
-2 \log^2 m_t+3 \log^2(m_t^2-t)
    \right. \NO \\
&&
+3 \log^2(m_t^2-u)
-2 \log(m_t^2-t)\log(t) -2 \log(m_t^2-u)\log(u)
-3 \log^2|m_{H^\pm}^2-m_t^2| \NO \\
&&
+4 \log (m_{H^\pm}) \log\frac{m_{H^\pm}^2-m_t^2}{m_t^2}
-4 \log (m_t) \log\frac{({{m_t}}^2 -t)({{m_t}}^2 -u)}{
tu(m_{H^\pm}^2-m_t^2)} 
-2 li_2 \bigg(\frac{m_{H^\pm}^2 s}{(m_{H^\pm}^2-t)(m_{H^\pm}^2-u)}
\bigg)
\NO \\
&&
-2 li_2 \bigg( \frac{( m_t^2 -t) u }{m_t^2 (u -m_{H^\pm}^2)} \bigg)
-2 li_2 \bigg( \frac{{{m_t}}^2}{{{m_t}}^2-{{m_{H^\pm}}}^2 } \bigg)
+2 li_2 \bigg(\frac{u-m_{H^\pm}^2}{ {{m_t}}^2-t} \bigg)
+2 li_2 \bigg( \frac{{{m_t}}^2}{ {{m_t}}^2-t} \bigg)
  \NO \\
&&
+2 li_2 \bigg(\frac{t-m_{H^\pm}^2}{m_t^2-u} \bigg)
-2 li_2 \bigg( 
\frac{ (t-{{m_{H^\pm}}}^2)(u-{{m_{H^\pm}}}^2) }{(m_t^2-t)(m_t^2-u)} \bigg)
+2 li_2 \bigg( \frac{{{m_t}}^2}{ {{m_t}}^2-u} \bigg)
-2 li_2 \bigg( \frac{t ( u-m_t^2 )}{m_t^2 (m_{H^\pm}^2-t)} \bigg)
  \NO \\
&&
+2 li_2 \bigg( \frac{{{m_{H^\pm}}}^2 (m_t^2-t) (m_t^2-u)
}{m_t^2(m_{H^\pm}^2-u) (m_{H^\pm}^2-t)} \bigg)
+2 li_2 \bigg(\frac{ts}{(m_{H^\pm}^2-t)(t-m_t^2)} \bigg)
  \NO \\
&&\left.
-2 li_2 \bigg(\frac{ m_t^2 s}{(m_t^2-t)(m_t^2-u)}  \bigg)
+2 li_2 \bigg(\frac{s u}{(m_{H^\pm}^2-u)(u-m_t^2)} \bigg)
\right\}.
\eea

The coefficients $d_2, d_1$ and $d_0$ of $D_0^2$ are 
 \bea d_2 &=& \frac{3}{2\,s \,
    \left( t - {{m_t}}^2 \right) },
\eea
\bea
d_1 &=&
\frac{1}{s \left( t - {{m_t}}^2 \right)}
\log\frac{m_t (m_{H^\pm}^2-m_t^2)}{s ({{m_t}}^2-t)^2},
\eea
\bea
d_0 &=&
\frac{1}{2 s \left(  {{m_t}}^2 -t \right)}
\left\{
\pi^2 \left[1-\theta(m_{H^\pm}-m_t)\right]+
2 \log^2(m_t)+ \log^2|m_{H^\pm}^2-m_t^2|-
\log^2(m_t^2-t) \right. \nonumber \\
&&-2 \log(m_{H^\pm}^2-t)\log(m_t^2-t)+4 \log\frac{s}{m_{H^\pm}^2-t}
\log\frac{m_t}{m_t^2-t}
-2 li_2 \bigg( \frac{m_{H^\pm}^2}{m_{H^\pm}^2-m_t^2}\bigg)
\NO \\
&&
+2 li_2 \bigg( \frac{m_t^2-t}{m_t^2} \bigg)
-2 li_2 \bigg( \frac{m_{H^\pm}^2 (m_t^2-t)}{m_t^2 (
m_{H^\pm}^2-t)} \bigg)+
2 li_2 \bigg( \frac{m_{H^\pm}^2}{m_{H^\pm}^2-t} \bigg)
\left.
+2 li_2 \bigg(\frac{m_{H^\pm}^2-t}{m_t^2-t} \bigg)
\right\}.
\eea

The coefficients $c_2, c_1$ and $c_0$ of $C_0^1$ are
 \bea c_2 &=& -\frac{1}{s}, \eea \bea c_1 &=&
\frac{\log(s)}{s}, \eea \bea c_0 &=& \frac{1}{6 s}\left[4 \pi^2 -3
\log^2(s) \right]. \eea

The coefficients $c_1$ and $c_0$ of $C_0^3$ are
 \bea c_1 &=& \frac{1}{m_{H^\pm}^2-t}
\log\left(\frac{ m_t^2-m_{H^\pm}^2}{m_t^2-t} \right), \eea \bea
c_0 &=&\frac{1}{2 (m_{H^\pm}^2-t)} \left\{\log^2 (m_t^2-t) -
\log^2(m_t^2-m_{H^\pm}^2)
+2 li_2\bigg(\frac{m_{H^\pm}^2}{m_{H^\pm}^2-m_t^2}  \bigg) -2 li_2
\bigg(\frac{t}{t-m_t^2}  \bigg) \right\}. \eea

The coefficients $c_2, c_1$ and $c_0$ of $C_0^6$ are
\bea c_2 &=& \frac{1}{2 (m_t^2-t)}, \eea \bea c_1
&=& \frac{1}{ (m_t^2-t)} \log\frac{m_t}{m_t^2-t}, 
\eea \bea c_0 = \frac{1}{2 (m_t^2-t)} && \left\{
-\log(m_t)+ \log^2(m_t^2-t) 
-2 li_2 \bigg(\frac{t}{t-m_t^2} \bigg) \right\}. \eea

$C_0^4$, $C_0^8$ and $D_0^3$ can be obtained from $C_0^3$, $C_0^6$ and $D_0^2$ by 
replacing $t$ with $u$.

\newpage
\begin{figure}
\epsfxsize=13 cm \centerline{\epsffile{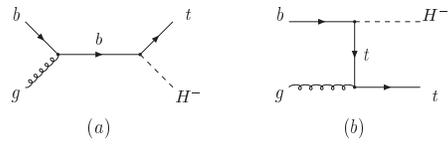}}
\caption[]{
Feynman diagrams at LO for $bg\rightarrow t H^-$.}
\end{figure}

\newpage
\begin{figure}
\epsfxsize=13 cm \centerline{\epsffile{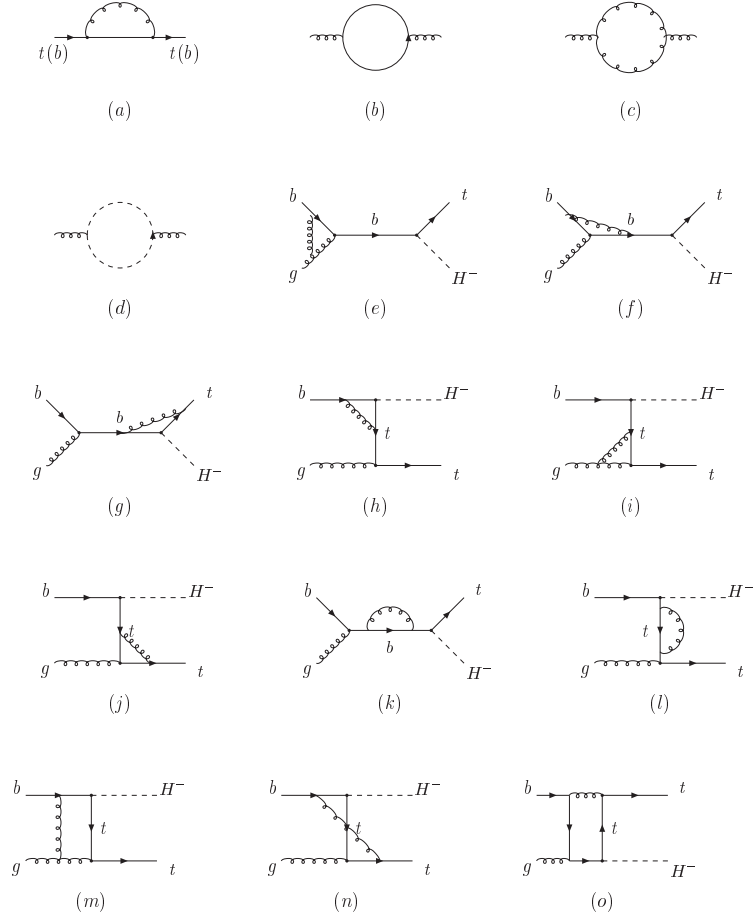}}
\caption[]{
Feynman diagrams of the virtual correction for the process $bg\rightarrow t H^-$.}
\end{figure}


\begin{figure}
\epsfxsize=13 cm \centerline{\epsffile{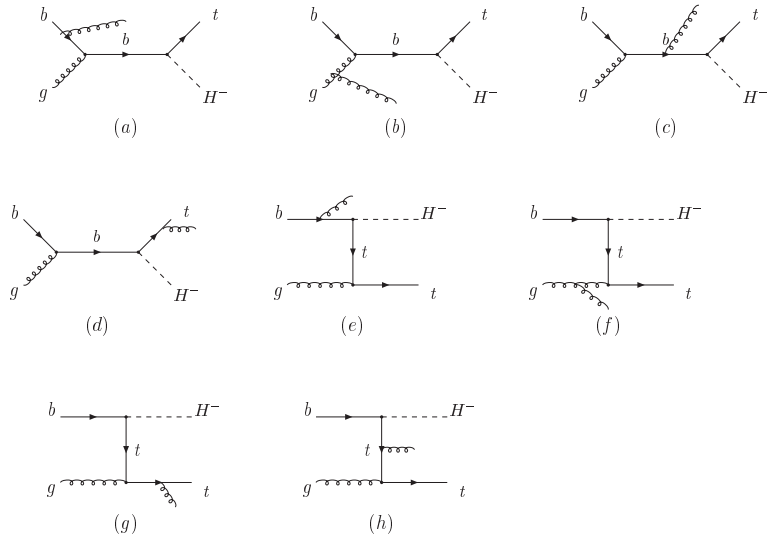}}
\caption[]{
Feynman diagrams of gluon-radiation process of $bg\rightarrow t H^- g$.}
\end{figure}

\begin{figure}
\epsfxsize=13 cm \centerline{\epsffile{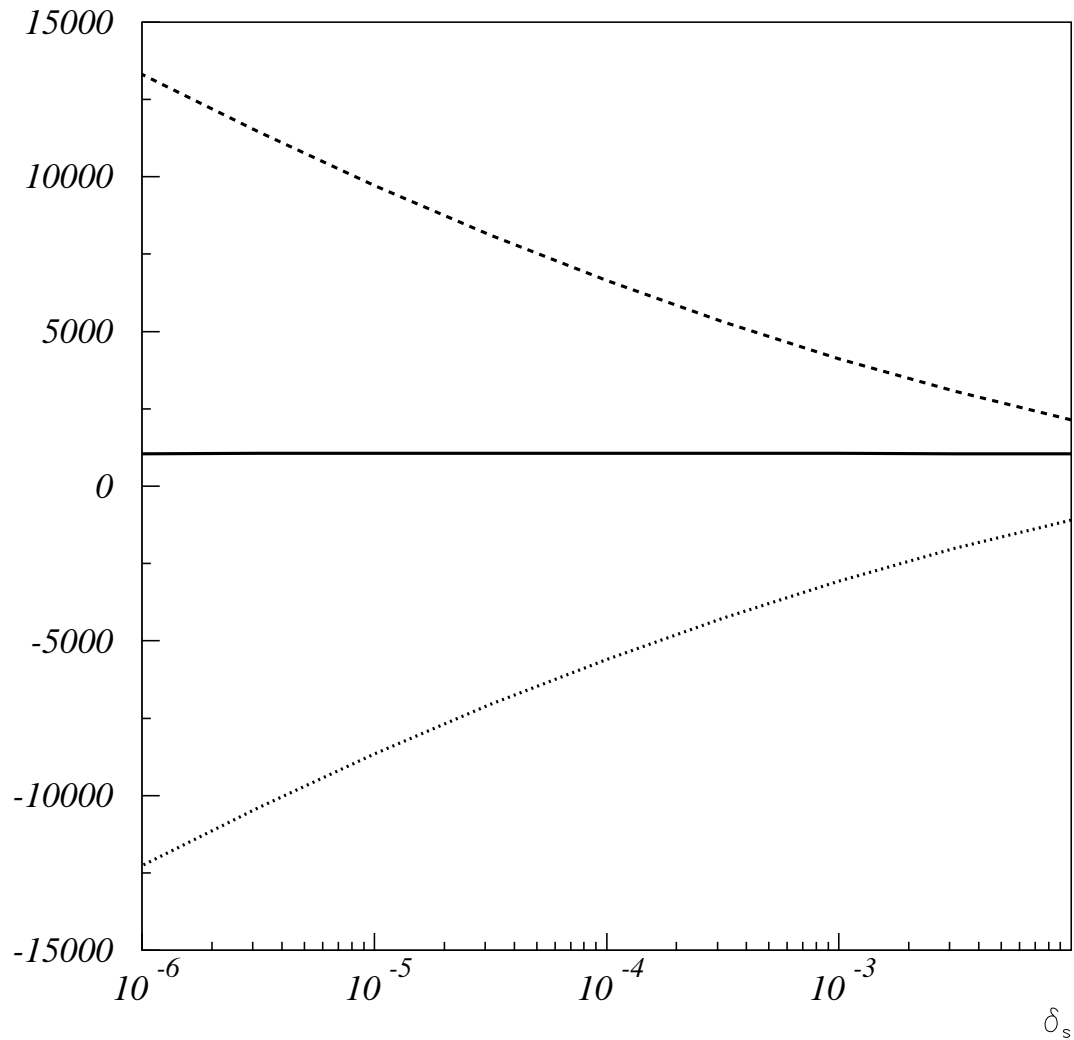}} \caption[]{
The cross sections [in $fb$] from hard non-collinear regions (dashed), other than hard 
non-collinear regions (dotted) and total (solid) as a function of $\delta_s$ with
$\delta_c=\delta_s/50$, where $\tan\beta=2$,  $m_{H^\pm}=200$ GeV and
renormalization and factorization scales $\mu=m_{H^\pm}+m_t$.}
\end{figure}

\begin{figure}
\epsfxsize=13 cm \centerline{\epsffile{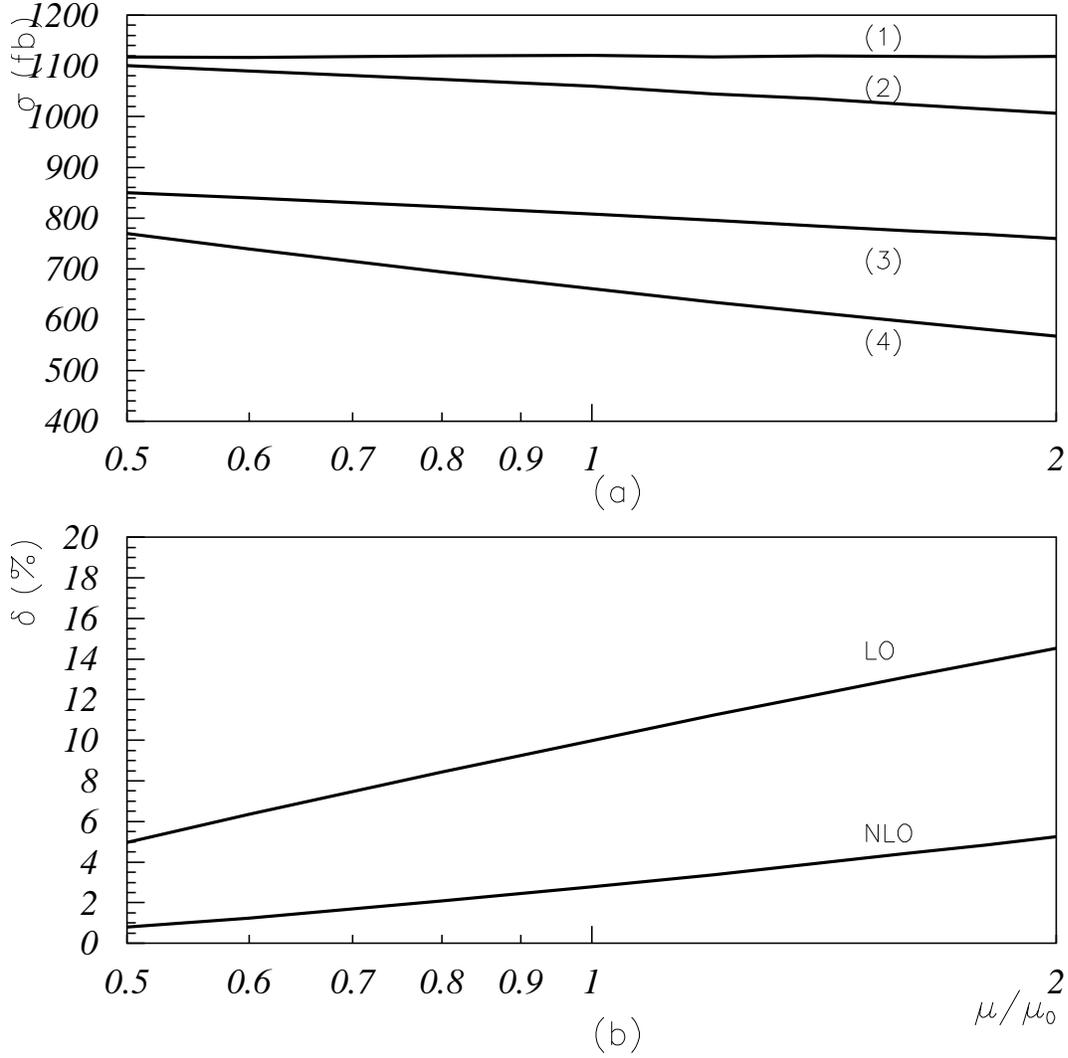}} \caption[]{
(a): the total cross sections as a function of $\mu/\mu_0$ with
$\mu_0=m_{H^\pm}+m_t$, where  $\tan\beta=2$ and $m_{H^\pm}=200$ GeV.
Curves (1)-(4) represent the cross section at NLO in OS 
scheme (1), NLO in  $\overline{MS}$ 
scheme (2), LO in OS
scheme (3),  LO in $\overline{MS}$ 
scheme (4).  (b): $\delta$ (defined in text) as a function of $\mu/\mu_0$.
}
\end{figure}

\begin{figure}
\epsfxsize=13 cm \centerline{\epsffile{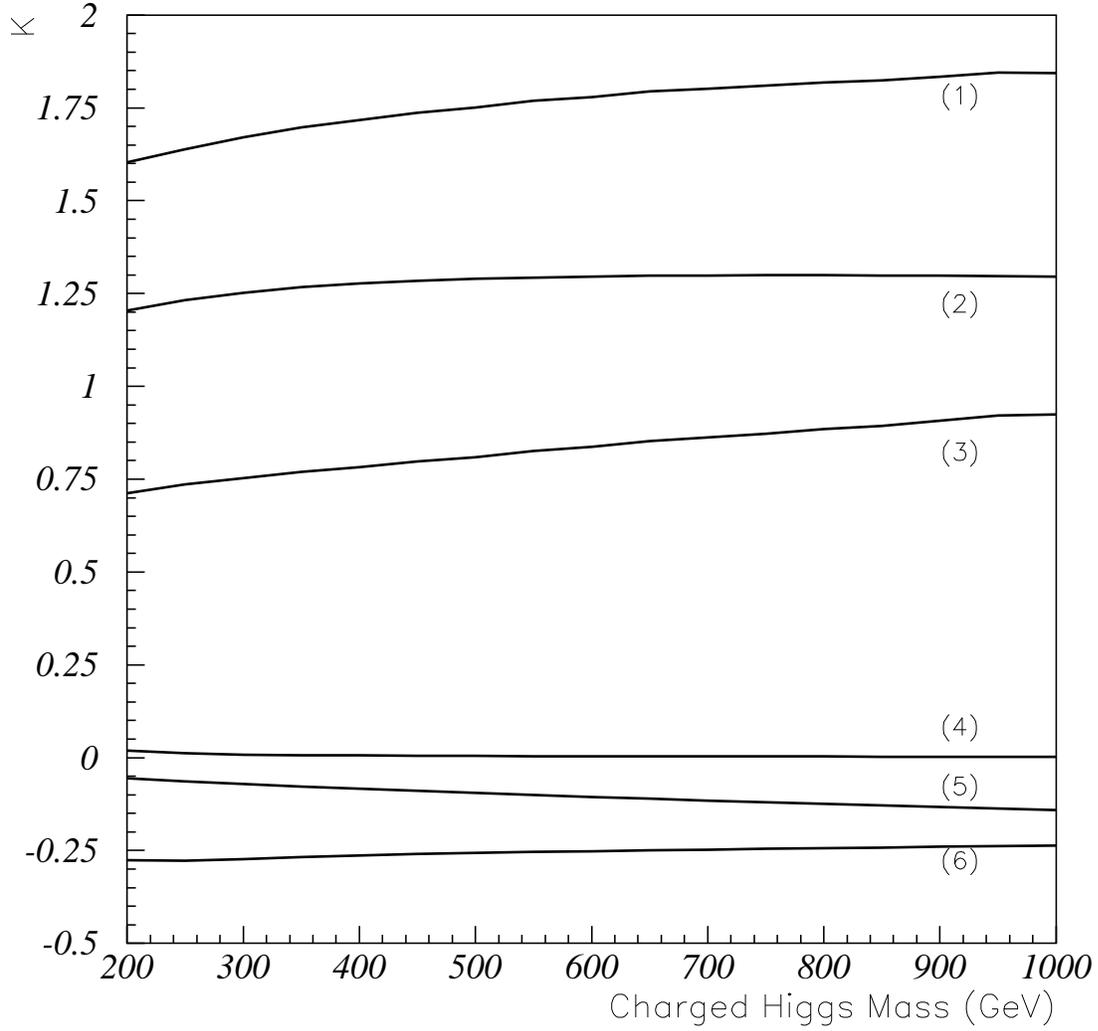}} \caption[]{
 Various contributions to K-factor versus $m_{H^\pm}$
with $\mu=\mu_0$.
Curves (1)-(6) represent contributions from all (1), improved Born (2),
virtual plus gluon-radiation (3), $q \bar q$ (4), $b q (\bar q)$ (5) and
initial-gluon (6).
}
\end{figure}

\begin{figure}
\epsfxsize=13 cm \centerline{\epsffile{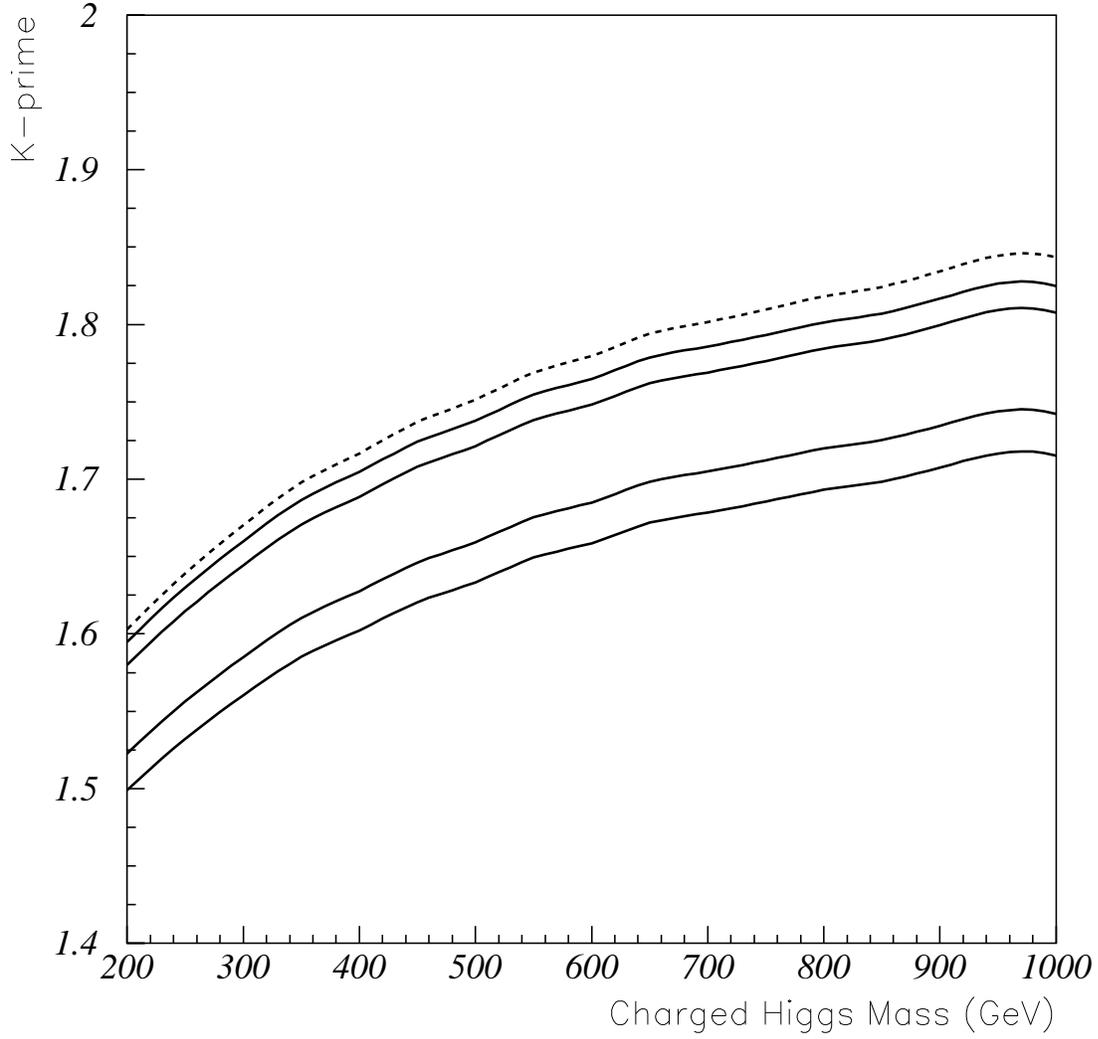}} \caption[]{
 $K^\prime$ (defined in text) as a function of charged Higgs boson
mass with $\tan\beta=2,5,10,30$ (solid lines from top to bottom) and
$\mu=\mu_0$. The dashed
line is K-factor in Fig.~6 which is re-shown for comparison. 
}
\end{figure}

\end{document}